\documentclass[aps,prl,reprint,showpacs,superscriptaddress]{revtex4-1}
\usepackage{graphicx}
\usepackage{dcolumn}
\usepackage{bm}
\usepackage{color}
\begin{document}


\title{The controllable super-high energetic electrons by external magnetic fields at relativistic laser-solid interactions in the presence of large scale pre-plasmas}

\author{D. Wu}
\affiliation{Shanghai Institute of Optics and Fine Mechanics, 
Chinese Academy of Sciences, Shanghai 201800, China}
\author{S. I. Krasheninnikov}
\affiliation{University of California-San Diego, La Jolla, California, 92093, USA}
\author{S. X. Luan}
\affiliation{Shanghai Institute of Optics and Fine Mechanics, 
Chinese Academy of Sciences, Shanghai 201800, China}
\author{W. Yu}
\affiliation{Shanghai Institute of Optics and Fine Mechanics, 
Chinese Academy of Sciences, Shanghai 201800, China}

\date{\today}

\begin{abstract}
The two stage electron acceleration model [arXiv: 1512.02411 and arXiv: 1512.07546] is extended to the study of laser magnetized-plasmas interactions at relativistic intensities and in the presence of large-scale preformed plasmas. It is shown that the cut-off electron kinetic energy is controllable by the external magnetic field strength and directions. Further studies indicate that for a right-hand circularly polarized laser (RH-CP) of intensity $10^{20}\ \text{W}/\text{cm}^2$ and pre-plasma scale length $10\ \mu\text{m}$, the cut-off electron kinetic energy can be as high as $500\ \text{MeV}$, when a homogeneous external magnetic field of exceeding $10000\ \text{T}$ (or $B=\omega_{c}/\omega_0>1$) is loaded along the laser propagation direction, which is a significant increase compared with that $120\ \text{MeV}$ without external magnetic field. A laser front sharpening mechanism is identified at relativistic laser magnetized-plasmas interactions with $B=\omega_{c}/\omega_0>1$, which is responsible for these super-high energetic electrons.

\end{abstract}


\maketitle

\textbf{\textit{Introduction--}}Recently, extremely strong magnetic field up to $1500\ \text{T}$ and with $\text{ns}$ duration is obtained by using a novel capacitor-coil target design that makes use of the hot electrons generated during the laser-target
interactions\cite{Sci.Rep.3.1170}. It can be expected that in the near future magnetic fields of still higher intensity can be realized. As we known that the dispersion relations or properties of laser-plasmas interactions with external magnetic fields significantly differ from that without magnetic fields, especially when the strength of the magnetic field is comparable to that of laser light, it is therefore of interest to investigate laser magnetized plasmas interactions with intense external magnetic fields, which might of great effects in both basic physics and practical applications\cite{PhyPla.15.056304,PhysRevLett.104.055002,PhysRevLett.100.225001,PhysRevLett.108.115004,PhyPla.20.023102,PRE.90.023101,Rev.M.Phys.85.751}, including particle acceleration, fast ignition in inertial confinement fusion, and intense radiation sources, et al.

In our previous work, we have studied the electron heating at relativistic laser solid interactions in the presence of large scale pre-formed plasmas\cite{PhyRevE.83.046401,PhyPla.19.060703,PhyPla.21.104510,arXiv1,arXiv2}. To explain the source of super high energetic electrons, we have proposed a two stage electron acceleration model\cite{arXiv1,arXiv2}. The first stage is the backward acceleration by the synergistic effects of charge separation electric field and pondermotive forces of reflected laser pulses. These fast electrons pre-accelerated in the first stage will build up electrostatic potential barrier with its peak energy several times larger than electron kinetic energy. Finally, electrons reflected by this potential barrier will acquire energy several times as large of their initial values.

In this work, external magnetic field is included to investigate the electron heating under relativistic laser magnetized plasmas interactions. We have found that electron energy spectra can be controlled by the external magnetic fields. Under the influences of external magnetic fields having $B=\omega_c/\omega_0>1$, the cut-off electron kinetic energy can be as high as $500\ \text{MeV}$, significantly higher than $120\ \text{MeV}$ of no external magnetic fields. The underlying physics of this dramatic energy enhancement is uncovered. We have identified a laser front sharpening mechanism at relativistic laser magnetized plasmas interactions regime with $B>1$, which is responsible for these super-high energetic electrons. 

\textbf{\textit{Simulation results--}}One-dimensional (1-D) particle-in-cell (PIC) simulation are run to model the processes of laser magnetized plasmas interactions. The simulation method is the same as our previous work\cite{arXiv1,arXiv2}. 
The size of the simulation box is $400\ \mu\text{m}$, with the region $0<z<100\ \mu\text{m}$ left as vacuum, and initial plasma density profile is taken as $n_e=n_{\text{solid}}/(1+\exp[-2(z-z_0)/L_p])$,  where $n_{\text{solid}}=50n_c$ is the solid plasma density, $z_0=180\ \mu\text{m}$ and $L_p=10\ \mu\text{m}$ is the pre-plasma scale-length. 
The laser of intensity $10^{20}\ \text{W}/\text{cm}^2$ enters the simulation box from the left boundary, and along the laser propagation direction, an uniform magnetic field is loaded. A diagnostic plane is placed at $z=300\ \mu\text{m}$ to temporally record the electrons passing through.

\begin{figure} 
\includegraphics[width=8.50cm]{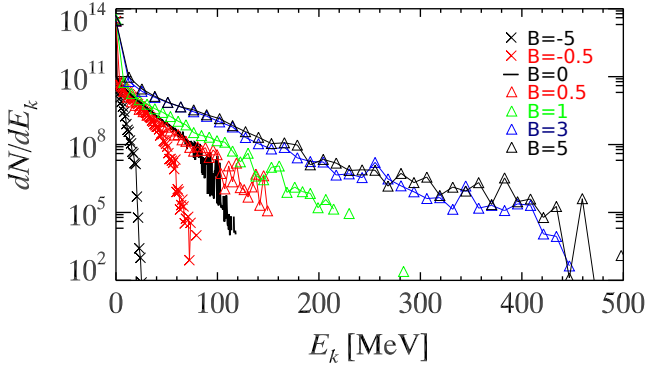}
\caption{\label{fig1} (color online) The final electron energy spectra recorded at $z=300\ \mu\text{m}$ under varying external magnetic fields, with pre-plasma scale-length $10\ \mu\text{m}$, laser wavelength $1\ \mu\text{m}$ and CP laser intensity $10^{20}\ \text{W}/\text{cm}^2$ fixed.}
\end{figure}

The incident laser is of right-hand circularly polarized (RH-CP), and the external magnetic field 
varying from $B=-5, -0.5$ to $B=0.5, 1, 3, 5$ is loaded to investigate the influence of external field on electron heating. The fast electron energy spectra obtained for varying external magnetic fields while keeping laser intensity $10^{20}\ \text{W}/\text{cm}^2$ fixed, are presented in Fig.\ \ref{fig1}. There is a clear relation between the electron cut-off kinetic energy and external magnetic field, as shown in Fig.\ \ref{fig1}. For external magnetic field with -z-direction, the electron generation efficiency is suppressed, i.e., the stronger of the external magnetic field the lower of the generation efficiency. On the contrary, for external magnetic field with z-direction, the electron generation efficiency is enhanced. Furthermore, we find that when $B=\omega_c/\omega_0>1$ ($\omega_c=eB/m_e$ and $\omega_0$ is laser frequency), the cut-off electron kinetic energy is dramatically increased from $120\ \text{MeV}$ to $500\ \text{MeV}$.    

\begin{figure*}
\includegraphics[width=16.50cm]{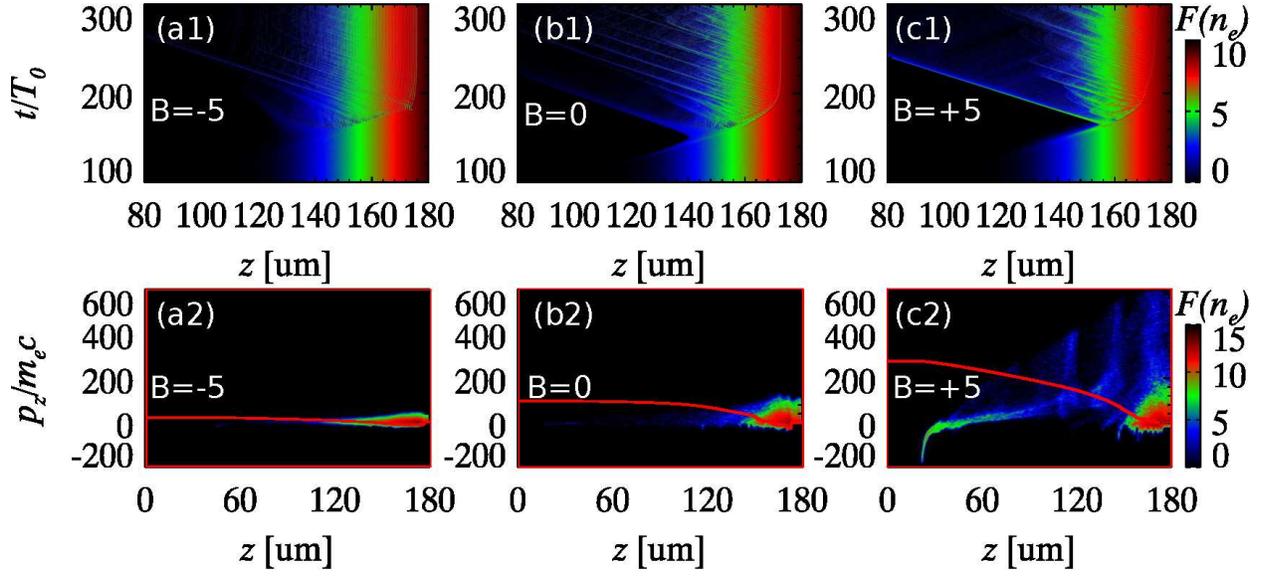}
\caption{\label{fig2} (color online) (a1)-(c1) Space-time evolution of electron density for different external magnetic fields. 
(a2)-(c2) The corresponding $z$-$v_z$ phase space plot of electrons at $t=300T_0$. The red curve covered on the phase plot is the electrostatic potential. The pre-plasma scale-length is of $10\ \mu\text{m}$, laser wavelength is of $1\ \mu\text{m}$ 
and the CP laser intensity is of $10^{20}\ \text{W}/\text{cm}^2$.}
\end{figure*}  

\textbf{\textit{Theoretical model--}}From the simulation results, we have found that electron energy spectra heavily depend on the strength and directions of the external magnetic fields, especially when condition $B=\omega_c/\omega_0>1$ is satisfied.
In Fig.\ \ref{fig2}, we have presented the space-time evolution and phase-space plots of electrons under three typical external magnetic fields, $B=-5$, $B=0$ and $B=5$. Compared with Fig.\ \ref{fig2} (a) and (b), from Fig.\ \ref{fig2} (c1) and (c2), we can identify the source of these super-high energetic electrons, which is from the very first electron bunch initially accelerated backward by the charge separation electric field. Furthermore, the very first electron bunch as shown in Fig.\ \ref{fig2} (c1) and (c2) seems to be of highly collimated. To fully understand the formation of this highly energetic and highly collimated electron bunch, we need to refer to Maxwell equations and the equations of motion of electrons at relativistic laser magnetized plasmas interaction regime.

The relevant Maxwell equations are
\begin{eqnarray}
&& \nabla\times\bm{E}=-\frac{1}{c}\frac{\partial \bm{B}}{\partial t} \\
&& \nabla\times\bm{B}=\frac{4\pi}{c}\bm{J}+\frac{1}{c}\frac{\partial \bm{E}}{\partial t}.
\end{eqnarray}

Assuming laser of the form $\exp[i(\bm{k}\bm{z})-i\omega t]$ and $\bm{J}=en_e\bm{p_e}/\gamma$, we can find
\begin{equation}
\label{p_E}
\bm{p_e}=\frac{-i\gamma(\omega^2-c^2k^2)}{4\pi e n_e\omega}\bm{E}.
\end{equation}

The electron's equations of motions are
\begin{eqnarray}
\label{eom}
&& m_e\frac{\partial {p_x}}{\partial t}=-e E_x-{ep_y B}/{\gamma} \\
&& m_e\frac{\partial {p_y}}{\partial t}=-e E_y+{ep_x B}/{\gamma}.
\end{eqnarray}

Combining Eq.\ (\ref{p_E}) into Eq.\ (\ref{eom}) and (5), we can easily obtain the dispersion relation of CP laser under magnetized plasmas
\begin{equation}
\label{dispersion}
k^2=\frac{\omega^2}{c^2}[1-\frac{\omega_{pe}^2}{\omega^2(\gamma-\omega_c/\omega)}],
\end{equation}
where $\omega_c=eB/m_e$ and $\gamma=(1+a^2)^{1/2}$ with $a=eE/m_e c$. According to Eq.\ (\ref{dispersion}), we find that for $\omega_c/\omega>1$, there is a resonant region with $\omega_c/\omega<\gamma<\omega_c/\omega+\omega_{pe}^2/\omega^2$. Within the resonant region, electromagnetic waves can not propagate. Instead, they are absorbed by plasmas. However for $0<\omega_c/\omega<1$ or $\omega_c/\omega<0$ (i.e. magnetic fields with -z-directions), there is no stopping band for relativistic intense laser beams.

\begin{figure}
\includegraphics[width=8.0cm]{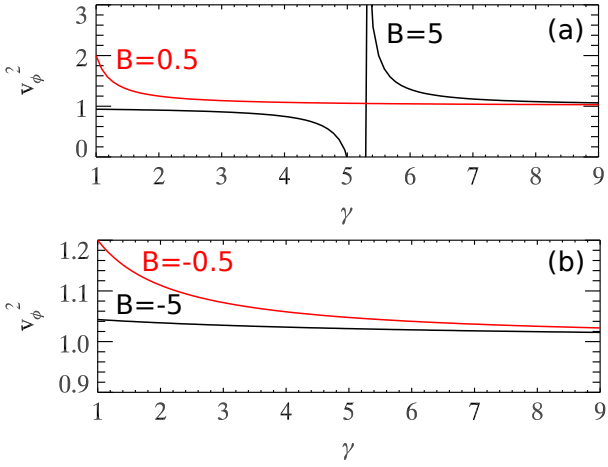}
\caption{\label{fig3} (color online) The $v_{\phi}^2$-$\gamma$ diagrams of relativistic intense electromagnetic wave in magnetized plasmas. Here we set $\omega_{pe}=0.5$ and $\omega=\omega_0=1$. (a) presents the cases with $B=\omega_c/\omega=5$ (black line) and $B=\omega_c/\omega=0.5$ (red line). (b) is the corresponding cases with $B$ of -z-directions.}
\end{figure}

Fig.\ \ref{fig3} shows the $v_{\phi}^2$-$\gamma$ diagram following the dispersion relation in Eq.\ (\ref{dispersion}), where $v_{\phi}^2=\omega^2/k^2c^2$. From Fig.\ \ref{fig3} (a), we can see that for external magnetic field of $B=5$, 
there is a resonant region with $5<\gamma<5.25$. 
Beyond this resonant region and with the increase of $\gamma$, we find $v_{\phi}^2$ is rapidly decreasing. 
For $\gamma\gg1$, we have $v_{g}\sim c^2/v_{\phi}$, which means $v_{g}$ is rapidly increasing. Considering a Gaussian laser propagating through the magnetized plasmas, 
these two effects, ``the resonant band'' and ``group velocity speeding with the increase of $\gamma$'', would lead to the laser front sharpening. As the ponderomotive force is determined by $\partial \gamma/\partial z$, the laser front sharpening effect would dramatically enhance the ponderomotive force of the laser beam. Compared with Fig.\ \ref{fig3} (b), it is clearly indicated that the laser front sharpening effect can be triggered only for $\omega_c/\omega>1$ and with the direction of external magnetic field in z-direction. 

\begin{figure}
\includegraphics[width=8.50cm]{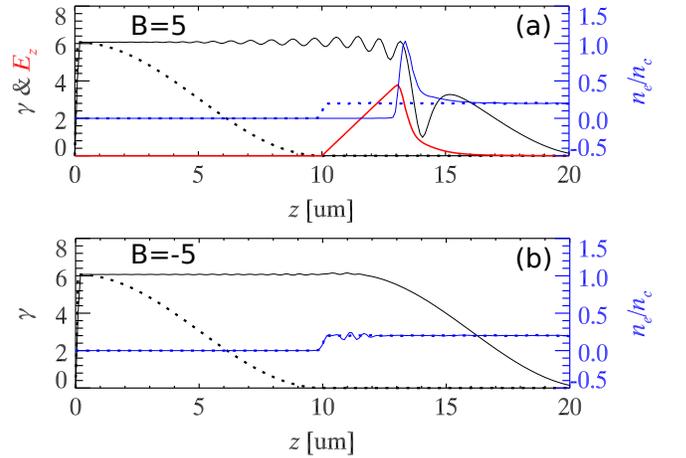}
\caption{\label{fig4} (color online) The uniform plasma is of constant density $n_e=0.2n_c$. 
The RH-CP laser is of intensity $10^{20}\ \text{W}/\text{cm}^2$, and the laser front is of $\sin^2$ profile 
with the rising period of $10T_0$. (a) and (b) show the laser amplitude and electron density profile with external magnetic field of $B=5$ and $B=-5$ at $t=10T_0$ (dashed) and $t=20T_0$ (solid), respectively.}
\end{figure}

To compensate the analysis above, we also run a serial of PIC simulations with constant background density $n_e=0.2n_c$ and $\sin^2$ rising profile having the rising period of $10T_0$. The laser is of RH-CP with intensity $10^{20}\ \text{W}/\text{cm}^2$. We change the directions and strength of the external magnetic fields to study their influences on the evolution of the laser front. 
As shown in Fig.\ \ref{fig4} (a),
for external magnetic field with $B=5$, there are two effects undertaken for the laser front. Within the region having laser amplitude $a\sim\gamma\sim5$, the electromagnetic waves are absorbed by plasmas due to the resonant motion of electrons. Beyond the absorbed region, we can also observe significant laser front congestion. The two effects, shown in Fig.\ \ref{fig4} (a), which collectively lead to the laser front sharpening, are well consistent with our theoretical analysis. In contrast, as shown in Fig.\ \ref{fig4} (b), if the external magnetic is of -z-direction, we do not observe the laser front sharpening effect, as expected by the theoretical analysis. 

From Fig.\ \ref{fig4} (a), at the resonant region $a\sim\gamma\sim5$, where the laser is absorbed by plasmas, we find intense electron accumulation therein. These accumulated electrons should be the source of the highly energetic and highly collimated electron bunch, as shown in Fig.\ \ref{fig2} (c1) and (c2). The laser front sharpening effect would dramatically increase the laser ponderomotive force, because of $f_{p}\sim\partial \gamma/\partial z$. This ponderomotive force would push all electron forward, leaving ions behind, until at a position where the charge separation induced electric force therein is equal to the ponderomotive force, i.e. $E_z\sim f_p$. As we know $E_z\sim n_e z$, we can easily conclude that the reflection position of the very first electron bunch depends on the external magnetic field, which is clearly confirmed by Fig.\ \ref{fig2}. For $B=5$, the reflection position is at $z=148\ \mu\text{m}$, while for $B=-5$, it is around $z=130\ \mu\text{m}$. 

The electron bunch will eventually be accelerated backward by the charge separation electric field. During this process, there exist a phase-locked mechanism which would result in the formation of highly collimated electron bunch. The profile of the charge separation field is shown with red line in Fig.\ \ref{fig4} (a). For electrons travelling faster, the electric field experienced by them is small, while for electrons travelling slower, the electric field acted on them is large. This kind of charge separation electric field profile will ``phase-lock'' the electrons, which share the same physics with the phase stable acceleration of ions by laser radiation pressure\cite{PhysRevLett.100.135003}, eventually leading to the formation of highly collimated electron bunch, as shown in Fig.\ \ref{fig2} (c1) and (c2).      

The highly energetic and highly collimated electron bunch would build an intense electrostatic potential, 
as shown by the red curves in Fig.\ \ref{fig2}, with its peak energy several times larger than electron kinetic energy. 
Finally, electrons reflected by this potential barrier will acquire energy several times as large of their initial values.

\textbf{\textit{Discussions--}}We already know the laser front sharpening effect, which is responsible for the dramatic enhancement of electron energy, only takes place for magnetic field with $B=\omega_c/\omega_0>1$. For Rd glass laser with typical wavelength $1.0\ \mu\text{m}$, the corresponding magnetic field can be as high as $10000\ \text{T}$. Although the highest static magnetic field obtained in experiment is $1500\ \text{T}$, the laser front sharpening effect and the related phoneme we reported in this work can still be confirmed by experiment. We would recommend to use CO$_2$ laser facility\cite{Nature.phys.8.95}. 
The laser wavelength of CO$_2$ laser is $10\ \mu\text{m}$, 
which means the $B$ can be as high as $1.5$, considering the highest static magnetic field reported so far. 

\textbf{\textit{Conclusions--}}The external magnetic field is included to investigate the electron heating under relativistic laser magnetized plasmas interactions. It is shown that the cut-off electron kinetic energy is controllable by the external magnetic field strength and directions. For external magnetic field with -z-direction, the electron generation efficiency is suppressed, i.e., the stronger of the external magnetic field the lower of the generation efficiency. On the contrary, for external magnetic field with z-direction, the electron generation efficiency is enhanced. Further studies indicate that for a RH-CP laser of intensity $10^{20}\ \text{W}/\text{cm}^2$ and pre-plasma scale length $10\ \mu\text{m}$, the cut-off electron kinetic energy can be as high as $500\ \text{MeV}$, when a homogeneous external magnetic field of exceeding $10000\ \text{T}$ (or $B=\omega_{c}/\omega_0>1$) is loaded along the laser propagation direction, which is a significant increase compared with that $120\ \text{MeV}$ without external magnetic field. The underlying physics of this dramatic energy enhancement is uncovered. A laser front sharpening mechanism is identified at relativistic laser magnetized plasmas interactions regime with $B>1$, which is responsible for these super-high energetic electrons. 

This work can be confirmed in experiment by CO$_2$ laser facility, whose wavelength is $10\ \mu\text{m}$, with the help of the highest static magnetic field, i.e. $1500\ \text{T}$, which is already obtained in experiment. Multi-dimensional effects shall be presented in a following separated paper. 

\begin{acknowledgments}
This work was supported by the National Natural Science Foundation of China (11304331, 11174303, 61221064), the National Basic Research Program of China (2013CBA01504, 2011CB808104) and USDOE Grant DENA0001858 at UCSD.
\end{acknowledgments}

{}

\end{document}